\RequirePackage{fix-cm}

\documentclass[10pt,journal,compsoc]{IEEEtran}

\ifCLASSOPTIONcompsoc
  \usepackage[nocompress]{cite}
\else
  \usepackage{cite}
\fi

\usepackage{graphicx}
\usepackage{algorithmicx,algpseudocode}
\usepackage{algorithm}

\algblockdefx{MRepeat}{EndRepeat}{\textbf{repeat}}{}
\algnotext{EndRepeat}

\begin{document}

\title{A generalized framework for the estimation of edge infection probabilities 
}


\author{Andr\'as B\'ota, Lauren Gardner
\IEEEcompsocitemizethanks{\IEEEcompsocthanksitem A. B\'ota and L. Gardner are with the Research Centre for Integrated Transport Innovation, 
School of Civil and Environmental Engineering, University of New South Wales, Sydney NSW 2052, 
Australia, e-mail: a.bota@unsw.edu.au, l.gardner@unsw.edu.au }}




\IEEEtitleabstractindextext{%
\begin{abstract}
Modeling the spread of infections on networks is a well-studied and important field of research. Most infection and diffusion models require a real value or probability on the edges of the network as an input, but this is rarely available in real-life applications. Our goal in this paper is to develop a general framework for this task. The general model works with the most widely used infection models and is able to handle an arbitrary number of observations on such processes. The model is defined as a general optimization task and a Particle Swarm heuristic is proposed to solve it. We evaluate the accuracy and speed of the proposed method on a high variety of realistic infection scenarios.
\end{abstract}

}

\maketitle



\section{Introduction}

\IEEEPARstart{I}{}nfection models are frequently used in many real-life applications in sociology, economics and epidemics \cite{filler1}.
There is a large variety of models available for different applications, the most
popular ones being the SI, SIR and SEIR models and their variants \cite{anderson,DIE}.
They are regularly used to model infectious diseases \cite{DIE}, the spread of information and behavior on social networks \cite{Granovetter}, 
or the spread of economic events \cite{DR,cejor,CSKKPT}. Some of the tasks associated with these models include influence maximization, that is 
finding the set of individuals yielding the largest expected infection \cite{KKT_1,filler2,filler3}, or the prediction of a posteriori infection values i.e. finding the probability of infection for all nodes of the 
network \cite{kinai2}.
Most infection models are network-based; they assume a set of nodes and the connections between them to be given. Most also assume
a specific value to be available on the links of the networks; these are called the {\em edge infection values} and they represent the probability that the infection spreads from one node to
another.

A common challenge in the application of infection models is the lack of information on the edge infection values on the links of networks.
Recently several authors have proposed models to estimate these values.  Many of the models assume that the time stamps of the infection for each node are given \cite{lg1,lg3,lg5,gomez,goyal,kimura,myers}, although some
approaches do not require this property \cite{spanyol,iip,amin}. The current estimation methods are based on a variety of infection models \cite{lg1}, 
but most often the SIR model is used \cite{gomez,goyal,kimura,amin}.
Most of these models were developed independently of each other and most of them take a different, often application specific approach to the problem.

Our goal in this paper is to define and solve a {\em general} model for the estimation of edge infection values of infection processes.
We denote this task as the {\em inverse infection task} and we propose the General Inverse Infection Model (GIIM) to solve such tasks. There are four main features of the GIIM that distinguish it from previous inverse infection modeling approaches. The features are listed below, and subsequently explained in further details.

\begin{enumerate}

\item The GIIM can account for an arbitrary number of observations on the infection process.

\item The GIIM allows for flexibility in the type of observations provided as input.

\item The GIIM allows for flexibility in the underlying infection model.

\item The GIIM allows for flexibility in the type of edge estimation values.

\end{enumerate}

An infection spreading process may be fully or partially observable, and some approaches even consider it to be non-observable. GIIM is able to handle an {\em arbitrary number of observations} on such processes, enabling it to adapt to the specific requirements of applications. In this paper we experiment with the two extremes of this requirement. In fully observed scenarios we know at exactly what time a node was infected. In two-observation cases we can only observe who was infected at the beginning and at the end. Both fully observed and two-observation approaches and any in between them can be naturally formulated in the framework of GIIM. 

Observations on the infection process might be available in various forms, so the model should be flexible enough to handle {\em different forms of input data}. Two types of observations are investigated in this paper. The first one is based on the common assumption where at a certain time step we know which nodes were infected and which nodes were not. In the second type of observation we have real values indicating how likely it is for the nodes to become infected at a given time step. We propose additional types of observations that we plan to investigate in future work.

Another key contribution is the freedom to choose the underlying infection model. Most of the existing models are applicable for a single infection one, most often the Independent Cascade (IC) model \cite{KKT_1} or another variant of SIR. In contrast GIIM makes the inverse infection tasks {\em independent of the infection model}. We show that the IC, SI, SIR, SEIR models can be used within the GIIM framework, making GIIM compliant with most of the infection models used in literature. 

Lastly, the GIIM is able to directly estimate real edge infection values, as well as functional forms which include variables representative of vertex and edge attributes. 

In this paper we evaluate the features of GIIM highlighted above using a large variety of artificial infection scenarios. The scenarios differ by graph classes, size of the underlying graphs, size of the outbreak, type of infection models used and the type of performance evaluation. Additionally, both fully and partially observed scenarios are considered with real and binary observations. Our goal is to analyze the behavior of GIIM on the above examples in terms of general  tractability of the tasks and the accuracy of the predictions.

The rest of the paper is organized as follows. In Section 2 we review some basic graph theoretic concepts and define the used infection models. In the third section we introduce the General Inverse Infection Model. 
We examine the various types of observations used in the model, possible ways to measure the goodness of the results and describe the optimization method itself. We also discuss a special case of the GIIM. In Section 4 we evaluate the performance of a number of specific variations of the general model. First we examine the tractability of the task on trees and directed acyclic graphs, then we use a high variety of artificially generated infection scenarios and complex networks to test the accuracy of our model. Finally, the scalability of the model is tested on large complex networks.
We close our discussion with a short description of some features of GIIM, that are not investigated in this paper, present some of the challenges posed in this study
and discuss possible ways to further expand the functionality of our model.

\section{Definitions}

The infection processes described in this paper take place on graphs. We will denote graph $G$ as $G(V, E)$, were $V_G$ is the vertex and $E_G$ is the edge set of $G$.
By default we are going to consider these graphs to be connected and undirected, otherwise we will note it in the text. 
The infection models require a real value $w_e \in [0,1], e \in E$ to be present on all edges of the graph, these are known as the edge infection probabilities. We will denote the surjective assignment of edge infection probabilities to the edges as $W_G: E(G) \mapsto [0,1]$.
The task of inverse infection is the estimation of these values.

\subsection{Infection models}

Among the most frequently used infection models present in the literature are the IC, SI,  SIR and SEIR models. These processes are iterative; events take place in discrete time steps, and the models terminate in finite steps.

Discrete infection models work by assigning states to the vertices of the graph. Each vertex may only be in one state at a given time, and all vertices must be in a state at all times. 
The states of the above models are Susceptible (S), Exposed (E), Infected (I) and Removed (R).
 The exposed and infected states have discrete time periods attached to them, we denote these as $\tau_e$ and $\tau_i$. 
Infected nodes attempt to infect susceptible neighboring nodes according to the corresponding edge weight probability $w_e$. If successful, the newly infected nodes transition into an exposed state for $\tau_e$ iterations, after which they attempt to infect susceptible neighbors for $\tau_i$ iterations, after which point they become removed (dead or immune) and can no longer infect other nodes. The models differ in complexity, the most complex being the SEIR model, which has all of the above states; in the SIR model $\tau_e = 0$; in the IC model $\tau_e = 0$ and $\tau_i = 1$ and in the 
SI model $\tau_e = 0$ and $\tau_i \rightarrow \infty$. Correspondingly in the IC, SI and SIR models the nodes are never in an exposed state, and in the
SI model the nodes are never removed, \textit{i.e.,} they may attempt to infect neighboring nodes indefinitely. 

We illustrate the mechanics of the infection process on the most general infection model, SEIR.
Apart from the network itself, the infection process of the SEIR model has two inputs. The first being the assignment of edge infection probabilities
$W_G: E(G) \mapsto [0,1]$ for all edges. The second being the set of initially infected nodes $A_0 \subset V$. These nodes are considered to be infected at the start of the process.
Let $A_i \subseteq V$ be the set of nodes in the infected state in iteration $i$. Each node $u \in A_i$ tries to infect its susceptible neighbors $v$ 
according to $w_{u,v}$, if the attempt is successful $v$ will begin its exposed period. If more than one node is trying to infect $v$ in the same iteration, the attempts are
made independently of each other in an arbitrary order within the same iteration. Still in iteration $i$ the status of infected nodes change to removed
if the difference between $i$ and the time of
their infection becomes greater than $\tau_e+\tau_i$; exposed nodes become infected when the difference becomes greater than $\tau_e$. If $A_t = \emptyset$ the process
terminates in iteration $t$. In a finite network the process always terminates in a finite number of steps.

The other infection models behave similarly, with minor differences. The SEIR infection processes take longer to complete because the nodes have to spend time in an exposed state before becoming infectious. Similarly, higher $\tau_i$ corresponds to more infected nodes because there are more opportunities to infect neighboring nodes. Finally, in the SI model, given enough time, all vertices become infected with a probability of one, if there is an initially infected node in
every connected component.

\section{General inverse infection framework}

In order to estimate the edge infection probabilities in inverse infection tasks we propose and define the General Inverse Infection Model (GIIM). The goal of GIIM is to formulate the inverse infection task as a general optimization problem by defining the inputs and outputs of such tasks and describing
the relationship between the infection and inverse infection models. In this section we provide a formal description of GIIM, illustrate how various inverse infection tasks can be represented in it, and propose the Particle Swarm Optimization method from \cite{FPSO} to compute the edge infection probabilities.

We assume that the underlying graph of the infection task is known. We also assume that we have observations on an infection process taking place on the network. The observations on the infection process are given in the form of $\vec{o}_t \in O$, where $t \in T$ denotes a time stamp.
Each $\vec{o}_t$ assigns an observation on the infection process at time step $t$ to all nodes of the network. Set $O$ contains all observations and set $T$ contains all time stamps.
At least two observations are needed to provide a meaningful model, but otherwise the number of observations required is not specified.
The nature of these observations will be defined in Section 3.1.
Finally we assume that we know the specific type of infection process taking place on the network, although in Section 5 we will discuss what happens if we relax this assumption.

We define the inputs of GIIM as follows: we have an unweighted graph $G$, an infection model $\mathcal{I}$, the set of sample times $T$
 and  the observations on the infection process $\vec{o_t} \in O$ for all $t \in T$, where $O = Inf(G, W,  \mathcal{I}, T)$. 
We denote the unknown weight assignment to the edges of the graph as $W_G: E(G) \mapsto [0,1]$, and for now we assume $ \mathcal{I}$ to
 be any network based infection model from section 2.1.
 We can interpret $Inf$ as a procedure that makes observations at time steps $T$ on infection process $\mathcal{I}$ taking place on graph $G$ with edge weights
$w_e \in W$, $e \in E(G)$. We also define a difference function $d$ that compares two observations: $d(O_1,O_2)$. The way we count the difference depends on the form of observations, for example if they are numerical values a vector norm is a natural choice; more examples will be provided in Section 3.2.
The task of GIIM is to find an estimation $W'$ of $W$ so that $d(O,O')$ between $O$ and $O' = Inf(G, W',  \mathcal{I}, T)$ is as small as possible: {\em we are looking for an edge infection probability assignment, that best explains the observations of the original process.}

According to this, we define the General Inverse Infection Model:

\smallskip

\noindent {\bf  General Inverse Infection Model:} {\em Given an unweighted graph $G$, and infection model $\mathcal{I}$, the set of sample times $T$ and 
observations $O = Inf(G, W,  \mathcal{I}, T)$, we seek the edge infection probability assignment $W'$ such that the difference $d(O,O')$ between 
$O$ and $O' = Inf(G, W',  \mathcal{I}, T)$ is minimal.}
\smallskip

The definition above defines the inverse infection task as an optimization problem: the minimization of error between a reference and a computed observation. The optimization
task itself would be an iterative refinement of the computed observation. The solution procedure goes as follow: begin with an initial weight configuration $W'_0$, run the infection model, extract observations $O'$, compute
the error between $O$ and $O'$, then refine $W'$ and repeat the process until the error is less than an accuracy constant $a$ selected by the user. Algorithm 1 summarizes the main components of GIIM.
The specific search strategy we implement in this paper to refine $W'$ is the Particle Swarm Optimization
method \cite{FPSO}, but other methods may be selected if they follow the structure above. It should be noted that the search process can be easily implemented in a parallel way. Given multiple candidate weight configurations, multiple infection models can be computed independently of each
other on multiple threads. We discuss the optimization task in more detail in section 3.3.

\begin{algorithm}[tpb]
\caption{\bf Generalized Inverse Infection Model}
\begin{algorithmic}[1]
\State Inputs: $G$, $\mathcal{I}$,  $T$, $O$, $a$
\State Choose initial edge infection probability assignment $W'$
\MRepeat
\State Compute $O' = Inf(G, W',  \mathcal{I}, T)$
\State Compute $d(O,O')$
\If{$d(O,O') \leq a$}
\State {\bf return} $W'$
\Else
\State Choose new $W'$
\EndIf
\EndRepeat
\end{algorithmic}
\end{algorithm}

GIIM gives a general framework for inverse infection, with many of the finer details intended to be application specific. In this work our evaluation focuses on the general performance under three flexible properties of the model:
the availability of observations, the type of observations, and the type of underlying infection model.  Two different error functions are considered to quantify the model performance. We also discuss  possible applications for our model, and provide a short description of extensions of our model in Section 5.

\subsection{Observations}

In the definition above $O$ is defined
as a set of vectors $\vec{o}_t \in O$ for all $t \in T$, and each vector assigns an observation to all vertices of the graph at a specific time step $t$. Since $T$ represents the set of sample times
it is safe to assume that it can be ordered, thus the set of observations can also be ordered according to $T$.

The cardinality of $T$ is not specified, but at least two
time samples are needed to make a meaningful prediction for $W$. The upper limit is application dependent. In case of a discrete time infection model, $T$ usually contains natural numbers,
where the infection process starts at time $i_0 = 0$.
If the infection model is finite, i.e. it finishes at iteration $i_{max}$ it is meaningless to make observations after $t_{max} = i_{max}$ because the process does not change after $i_{max}$.  Thinking in the opposite way, the first and last observations -- at $t_0$ and $t_{max}$ respectively -- also give bounds, since we do not have other
observations apart from the ones at $T$. Therefore if we want to explain the witnessed infection scenario we should consider $t_0$ and $t_{max}$ to be the beginning and the end
of the infection process. It is possible to {\em predict} the behavior of an infection process beginning before $t_0$ or ending after $t_{max}$;
we will investigate this in future work.

The task of GIIM is to find an edge weight configuration that explains the original observed infection process. Therefore, the set of observations $O$ should contain information that represents the behavior of nodes in the infection process. In this paper we suggest two extreme cases of observation levels, but the proposed framework is able to handle other types as well, which we identify in Section 5.

\begin{enumerate}

\item Two observations are available at $t_0 = i_0$ and $t_1 = t_{max} = i_{max}$.

\item Observations for all iterations of the discrete infection process are available $T : \{t_0 = i_0, \dots, t_{max} =  i_{max}\}$.

\end{enumerate}

Realistic cases may be in between these two extremes, all of which can be accommodated within the modeling framework. We will investigate the relationship between the number of available observations and the accuracy of the estimation in Section 4.

The set of observations $O$ should contain information that represents the behavior of nodes in the infection process. In this paper we suggest two specific \textit{types} of observations that can be made; the proposed framework is able to handle other types as well, which we will discuss in Section 5.

The first observation type defines the observations as real values between $0$ and $1$, which are assigned to each node at the time steps where observations are available. For a time step $t$ and node $v$ the value  $o_{v_t}$ represents the likelihood that node $v$ is infected at time step $t$  i.e. the cumulative infection probability of $v$ up to time step $t$, while $\vec{o}_t$ represents the vector of observations for all nodes at time step $t$. The initial infections, the probability that $v \in V(G)$ is infected before the process, are stored in $\vec{o}_{t_0}$, while the probability that a node becomes infected measured at the end of the process are in $\vec{o}_{t_{max}}$, corresponding to the a priori and a posteriori distributions assumed known in the first case.  For finite infection models, if we limit our viewpoint to a single node, the series of observations, $o_{v_t}$ will be a monotonically increasing function since the $o_{v_t}$ starts at the initial infection probability $o_{v_0}$ and the infection process may never decrease this value.




Another way to make observations is to simply consider flags on the nodes signaling whether they have been infected or not. In this case, for a node $v \in V(G)$, $o_{v_t} = 0$ indicates it has not been infected up to time step $t$, and $o_{v_t} = 1$ means it is already infected at $t$. This approach is the most common in literature \cite{lg1,gomez,goyal}. It can be used directly, or if time stamps are available on nodes we can simply consider $o_{v_t} = 0$ if $t < t_{inf}$ and $o_{v_t} = 1$ if $t \geq t_{inf}$, where $ t_{inf}$ denotes the time stamp, \textit{e.g.}, the time at which a node was infected. We can construct a series of observations this way; in the worst case if every node has a different time stamp $|T| = |V(G)|$. 

The analysis in this paper is restricted to these two observation cases. 

\subsection{Error function}

The chosen error function should be able to compare observations made on the infection process. The observations are a set of vectors with time stamps attached to each vector. We can pair the vectors according to their time stamps, compute the error and then aggregate the individual distances into a single value. 

If the observations are vectors containing real numbers, a form of vector distance should be a natural choice.
For the continuous valued observation type presented in the previous subsection we use the root mean squared error (RMSE) function to compare the individual observation vectors. For binary observations ROC evaluation is used in  similar fashion.

\subsection{Computation}

The GIIM defines a general optimization task, which can be solved with the iterative refinement of an initial edge weight configuration. The number of edges even in a medium sized graph can be in the tens or hundreds of thousands, so finding the proper combination of weights can be challenging, therefore choosing an appropriate optimization method is crucial. Here we a recommend and use the Fully  Informed Particle Swarm method of Kennedy and Mendes \cite{PSO,FPSO}, which has proved successful in previous inverse infection tasks \cite{cejor}.

The optimization task was defined at the beginning of Section 3 as the minimization of the error between computed and reference observations. The method is initialized with randomly selected 
edge infection values within reasonable bounds, then a simulation of the infection process is computed to create observations. The error between the reference and computed
observations is counted and used to adjust the edge infection values. The process is then repeated.

The search strategy is Particle Swarm Optimization: an iterative multi-agent non-gradient metaheuristic. Each agent has a position that represents an edge weight configuration, 
and the agents refine their position by interacting with their neighbors. This is done by adding the velocity of the agent to the position of the agent. The velocity of the agent is computed from its previous value, the best result found by the agent and its neighbors and a random vector.
The goodness of an agent is the goodness of the edge weight configuration it represents, and the goodness is
evaluated by running an infection model and computing the error of the resulting observations. The agents are connected to each other through a pre-defined topology describing
the neighborhood of each agent.

The specifics of the update rules for the positions and velocities as well as the agent topology are not fixed, there are several approaches in the literature. Here, like in \cite{iip}, we follow the recommendations of Kennedy and Mendes \cite{FPSO} by adopting a von Neumann neighborhood for the agents and the following update rules:

\begin{equation}
\vec{v}_i \gets \chi \left(\vec{v}_i + \sum_{n =1}^{N_i}\frac{U(0,\varphi)(\vec{b}_{nbr(n)} - \vec{x}_i)}{N_i}\right),
\end{equation}
\begin{equation}
\vec{x}_i \gets \vec{x}_i + \vec{v}_i,
\end{equation}

\noindent where $\vec{x}_i$ and $\vec{v}_i$ denotes the coordinate and velocity of particle $i$, $U(min,max)$ is a uniform random number generator,
 $\vec{b}_i$ is the best location found so far
by particle $i$, $N_i$ is the number of neighbors $i$ has and $nbr(n)$ is the $n$th neighbor of $i$. The formula has two parameters: $\chi$ is the constriction coefficient and $\varphi$
is the acceleration constant. Again, we use the recommendations of Kennedy et al., and set $\chi = 0.7298$ and $\varphi = 4.1$.

\begin{algorithm}[tpb]
\caption{\bf Particle Swarm Optimization}
\begin{algorithmic}[1]
\ForAll{$a_i$}
\State Initialize $\vec{x}_i$ for agent $a_i$ within the boundaries of the search space
\State Initialize $\vec{v}_i$ for agent $a_i$
\State Set $\vec{b}_i \gets \vec{x}_i$
\State Select the neighbors of $a_i$ according to the topology
\EndFor
\Repeat
\ForAll{$a_i$}
\State Update $\vec{v}_i$ according to equation 2
\State Update $\vec{x}_i$ according to equation 3
\State Calculate the error function $e(\vec{x}_i)$ in position $\vec{x}_i$
\If{$e(\vec{x}_i) < e(\vec{b}_i)$}
\State $\vec{b}_i \gets \vec{x}_i$
\EndIf
\EndFor
\Until{termination criterium is met}
\end{algorithmic}
\end{algorithm}

\medskip

At the beginning, the position of the agents is initialized with random values, and the velocities are set to zero. Then at the end of each iteration the positions and velocities are updated
synchronously according to the above equations. The search is stopped if the method does not find a better global result in a set number of iterations. Algorithm 2 gives an outline of PSO.

The above algorithm can be implemented in a parallel way. Since the positions and velocities of each agent are updated at the end of each iteration,
the goodness of the current position of each agent can be computed independently and simultaneously on multiple threads. This way the running time of the algorithm can be decreased
significantly in multi-core processors.

\subsection{IIP as a special case of GIIM}

The Inverse Infection Problem (IIP) as appeared in \cite{spanyol,iip} can be considered the predecessor of GIIM.
GIIM uses the same FPSO optimization method as IIP, and the RMSE error function is one method used to guide the search in both models. However, compared with GIIM, the IIP is much more limited in terms of scope. In GIIM a range of observation levels can be accommodated, thus the error function has to be extended to account for multiple real vectors. Additionally, IIP fixes the number of observations to two: the a priori observation at time point zero, and the a posteriori observation after the infection process concludes. The nature of the observation is also fixed: IIP uses real infection values for the nodes: the probabilities of
infection before and after the infection process. The IIP is also only applicable for one type of infection model, the Generalized Cascade Model \cite{acta}, which is a variant of the IC model. In contrast, GIIM handles an arbitrary number of observations, does not constrain the type of these observations, and does not specify the used infection model.

\section{Evaluation}

The GIIM was designed to be a versatile tool, that is easy to adapt to real-life problems. As we have seen in the previous section it offers flexibility in terms of
input data (number and type of observations), error function and infection model. Investigating all possible combinations of these is beyond the scope of this paper, 
instead here we focus on properties best tested on artificial infection scenarios. We 
measure the difficulty of a variety of inverse infection tasks in terms of the accuracy and the speed of the estimation. The tasks vary based on the number and type of available
observations, the infection model and the prevalence of infections in the graph. A more detailed description of the tasks is provided in Section 4.3.

We present the evaluation in three parts. Section 4.2 focuses on the feasibility of the estimation on simple graph classes, namely trees and directed
acyclic graphs. We point out some trivial and non-trivial properties of the problem. 
In Section 4.3 we test the general applicability of the method on complex networks. Here we consider a number of infection scenarios with different transmission
probabilities between nodes, prevalence of initial infections, infection models and their parameters. We also examine how the number of available observations affect the quality of the
estimations. In Section 4.4 we consider the scalability of the method on large networks and the stability of the optimization method.

\subsection{Method details and experiment setup}

The {\em inverse infection tasks} in this paper are constructed according to the following procedure.

\begin{enumerate}

\item For a given graph, reference edge weights and the initially infected vertices are selected. The reference edge weights are not used in the inverse infection model, so they are omitted afterwards.

\item An infection model is selected and simulated using a modified version of CompleteSim published in \cite{acta}. The original method computes two observations for the IC model, but it is easy to adapt it to other models. The set of infection models considered includes:

\begin{enumerate}

\item The Independent Cascade Model (SIR with $\tau_i = 1$).

\item The SIR model with $\tau_i$ infectious period.

\item The SEIR model with $\tau_e$ latency and $\tau_i$ infectious period.

\item The SI model.

\end{enumerate}

\item During the simulation process, the set of reference observations for all nodes is computed. The observations are taken at the following time steps:

\begin{enumerate}

\item Two observations are available at $t_0 = i_0$ and $t_1 = t_{max} = i_{max}$.

\item Observations for all iterations of the discrete infection process are available $T : \{t_0 = i_0, \dots, t_{max} =  i_{max}\}$.

\end{enumerate}

\item Observations may be in the following form:

\begin{enumerate}

\item $o_{v_t} = p_{v_t}, t \in T$ for all $v \in V(G)$, that is {\em real infection values} for all vertices of the graph denoting the probability of infection up to time step $t$. $T$ may contain an arbitrary
number of observations with respect to the restrictions in section 3.1.

\item $o_{v_t} = b_{v_t}, t \in T$ for all $v \in V(G)$, that is {\em binary flags} for all nodes of the graph signaling that the node was in an infectious state at time step $t$. $T$ may contain an arbitrary
number of observations with respect to the restrictions in section 3.1. 

\end{enumerate}

\item An error function is attached to the task.

\begin{enumerate}

\item The RMSE function averaged over all real-valued observations.

\item The ROC AUC value for the last observation with binary observations.

\end{enumerate}

\item The Particle Swarm Optimization method computes the estimated edge infection probabilities and observations. 
The method is used with the number of agents ranging from 100 to 250 depending on the size of the task. The method is stopped if it does not find a better solution in
50 iterations. We experimented with other settings and found, that the accuracy of the method cannot be improved significantly by increasing the PSO parameters.

\end{enumerate}

The final error value measures the goodness of the estimation.
This procedure is repeated for a range of graph classes, sizes and infection scenarios.

\subsection{Acyclic examples}

We examine two specific graph classes in this section: trees and directed acyclic graphs to highlight certain non-trivial properties of the problem, but it is worth
noting that fully observed finite infection processes (with time stamps for each node) can be 
naturally represented as DAG-s.
Our goal here is to find out how difficult it is to compute inverse infection tasks for these seemingly simple graph classes. 

\subsubsection{Underdetermination}

A significant challenge exists in the estimation of edge infection values, which is illustrated using the following small example. Consider the example network in Figure 1/a.
There are two vertices, A and B, connected by a directed edge pointing from A to B. A and B are infected in succeeding iterations, A in $t=0$ and B in $t=1$, therefore both have an observed binary flag of one. 
For the sake of simplicity we consider
the IC infection model. The best and optimal way to explain this scenario is to consider the connecting edge to have an infection probability of one. This is
obviously an artifact rising from the available observations, a lack of information since we have only one observation on a highly stochastic process. From the optimization point of view however this is an optimal solution to the scenario, since this is the most likely explanation of the observations.
If we have real-valued observations on vertex A and B the edge infection probability is the one satisfying the equation $p_B = p_A*w_{AB}$.
Next consider the case shown in Figure 1/b. 
There are now two possible sources of infection for B. Vertex A and vertex C both have a directed edge pointing towards B and both were infected in the iteration before B was infected.
Suppose that $p_A$ and $p_C$ are independent of each other. Now $p_B = 1 - ((1 - p_A*w_{AB}) * (1 - p_C*w_{CB}))$. If the edge infection probabilities are real there are many, possibly infinite
ways of explaining this event, and this occurs at both binary flags and real $p_v$-s. The challenge in accurate edge weight estimation again arises from a lack of information: we cannot decide the value of one edge value without
knowing the other. Similar phenomena can be observed in the other infection models.

\begin{figure}[!t]
\label{fig:`}
\centering
\includegraphics[scale=0.6]{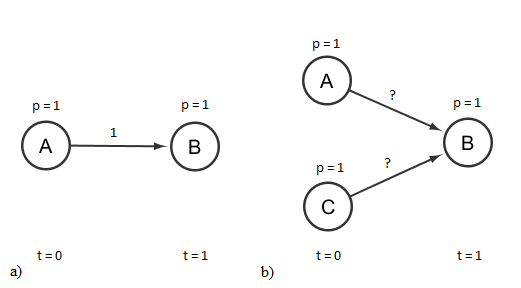}
\caption{Example of underdetermination. There is only one edge weight, that explains the infection event on a), while b) is underdetermined.}
\end{figure}

We will refer to the problems above as the {\em underdetermination} of the inverse infection task. 
Underdetermination occurs in most real-life examples, even when the infection process is completely observable. 


\subsubsection{Trees and Directed Acyclic Graphs (DAG-s)}

We first evaluate the performance of GIIM with arborescences.
In these graphs the infection spreads from the root, which is infected with probability of 1.  Reference edge values were drawn from a uniform distribution between 0 and 1.
Since there is only one directed path from the source of infection to each node, identifying the edge infection probabilities corresponding to the task should be easy. 
We consider both binary and real vertex infection values, for three binary trees with heights four, eight and eleven
and two DAG-s with 42 and 420 edges.

\begin{table}[!t]
\caption{Averaged error of the optimization method on trees and directed lattices of various sizes for the two-observation case.
The graph class, number of edges, the accuracy of the results (error at the end of the optimization process) can be seen. Both binary and real-valued observations were tested, with
AUC and RMSE provided for the former, only RMSE for the latter.
Result are provided for the IC infection model.}
\centering
\begin{tabular}{lllll}
\hline\noalign{\smallskip}
Class & Edges & Binary & Binary & Real-valued \\
 & & ROC AUC & RMSE & RMSE  \\
\noalign{\smallskip}\hline\noalign{\smallskip}
Tree & 31 & 1 & 0 & 0.0171 \\
Tree & 511 & 1 & 0 & 0.0269 \\
Tree & 2047 & 1 & 0 & 0.08  \\
DAG & 42 & 1 & 0 & 0.0262  \\
DAG & 420 & 1 & 0 & 0.0562  \\
\noalign{\smallskip}\hline
\end{tabular}
\end{table}

First consider the task where we only have two observations on the infection process, one at the beginning $t_0 = i_0$ and one at the end $t_1 = t_{max} = i_{max}$ of the 
process\footnote{We only consider the IC, SIR and SEIR models in this section.}.
The accuracy of the estimations are presented in Table 1. The differences between infection models was minimal, therefore results are only shown for the IC infection model.
When the observations are provided as binary flags the optimization algorithm was always able to find an optimal solution with RMSE = 0 and AUC = 1 much faster than with the real valued solution. The reason for this is due to the underdetermination explained
in the previous section. If the observations are vectors of ones and zeros, the best way to explain the observations is to assign the edge values to be ones and zeros. To further illustrate this concept, consider the example shown in Figure 2.
A leaf node in the tree has an observed infection value of one if all the edges leading to the leaf from the root have an edge infection value of one. If a leaf 
node has a value of zero, at least one edge connecting it to the root must have a zero value. This constrains the search to binary values as opposed to the continuous values of the
real-valued case. Even if the behavior of the infection process is more complicated in DAG-s, i.e. there are more explanations, this makes the binary inverse infection task much easier.

\begin{figure}[!t]
\label{fig:`}
\centering
\includegraphics[scale=0.6]{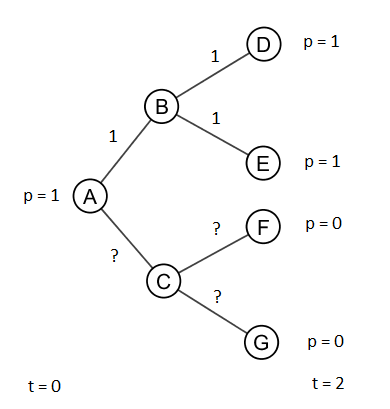}
\caption{Example of underdetermination. A leaf node has an observed infction value of one if and only if all the edges leading to it have edge infection values of one. A leaf node has an observed infection value of zero if at least one edge connecting the root to it has a value of zero.}
\end{figure}

Table 1 shows that in the two-observation real-valued case the second DAG with 420 edges is more challenging than the similar sized tree with 511 edges. This is because 
in the trees there is only one path from the source of the infection to each node, so there is
only one way to explain the process, while in the DAG-s multiple paths are present, meaning there are multiple optima. The first
problem is therefore easier than the second, hence the difference in the real-valued RMSE results in Table 1.

If the infection process is fully observed, \textit{i.e., } infection status at all timesteps is available, the error on all nodes for each individual time step can be computed separately, therefore it is possible to measure the performance of the optimization method in each iteration of the infection process.
Figure 3 shows results for the fully observed case for the second tree and the biggest DAG.
Since the root is the source of the infections and there are no directed cycles, an
observation at time step $i$ defines the infection values for nodes that are up to $i$ distance from the root. 
The error steadily increases as the infection reaches further from the root: at the last observation the RMSE value
is significantly greater than the value measured at the two information case. 
This increase in error is because the longer the chain of infections leading to a node, the more difficult it is to estimate its value. A possible reason for this is, the error becomes "spread out" equally among observations. In other words, the first few time steps are easier to compute because the nodes at a short distance from the root are far less numerous than the ones farther away. Additionally, the chain of edge infection values connecting them to the root is shorter, so there are less values to estimate.

\begin{figure*}[!t]
\label{fig:1}
\centering
\includegraphics[scale=0.6]{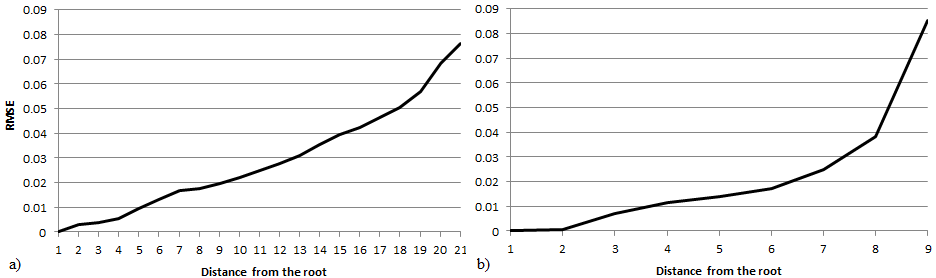}
\caption{The RMSE at different distances from the source in the fully observed IC model for a) the second DAG and b) the second tree in Table 1.}
\end{figure*}

From the pure optimization perspective, the task is somewhat difficult, since the number of parameters to estimate -- the number of edges -- can be large.
For the real-valued scenarios presented the total error was always below 9\%, and significantly lower on the smaller examples. For the binary case the method was able to find the best solution every time.
The PSO method required between 100 and 1500 iterations to find the above solutions, with the larger graphs taking more time and iterations, but
there was minimal difference between the accuracy of the results, further illustrating the stability of the chosen method. The SIR and SEIR models took slightly more time to compute than the IC due to the longer and more complex infection process (non-zero $\tau_i$ and $\tau_e$).

\subsection{Complex networks}

In addition to trees and DAG-s, we also examine the performance of GIIM on a more general complex network structure, and under several different infection scenarios.
The graph in question is undirected and it has 250 nodes and 897 edges. The graph was created with the forest fire model in \cite{forestfire}. The network itself is relatively small in network science terms, but our task here is to find the correct value for each edge, which is moderately challenging for the optimization algorithm. 
We examine the scalability of the method in the next section.

Our goal in this section is to compare the performance of the method in terms of accuracy and running time under a number of fundamentally different infection scenarios.
The scenarios differ according to the infection model, type and number of observations, the fraction of initially infected nodes and the size of the reference edge infection values.
The type and number of observations will be the same as in the previous section. We measure RMSE on real-valued observations and ROC AUC on binary ones, and
consider minimally and fully observed cases as defined in Section 4.1. The scenarios examined are summarized as follows:

\begin{itemize}

\item Four different infection models: IC, SI,  SIR with $\tau_i = 2$, SEIR with $\tau_e = 2$ and $\tau_i = 3$.

\item Three sets of initially infected vertices: 33\%, 10\% or 2\% of the nodes are initially infected with the infection probabilities ranging between 0 and 0.5 for the real-valued case, and uniformly 1 for the binary case.

\item Three sets of reference edge infection values as defined in Section 4.1. The values
were drawn from an uniform random distribution between 0 and 0.25, 0 and 0.5, 0 and 0.75 respectively.

\end{itemize} 

If we only consider the infection scenarios we have $4*3*3 = 36$ tasks, and if we add the observation types this goes up to $108$. Each of these tasks was computed ten times and
the results were averaged. The standard deviation measured between results was below 10\% for all infection scenarios.

\begin{table*}[!t]
\caption{Accuracy of the optimization method on complex networks for the two-observation case. RMSE is shown for the real-valued case, 
and AUC is shown for the binary one.
The numbers of the inverse infection tasks correspond to the scenarios described in this section. The first number denotes the percentage of initially infected nodes, the second one the maximum values of the edge infection probabilities.}
\centering
\begin{tabular}{lllllllll}
\hline\noalign{\smallskip}
Task  & IC &  & SIR 2 & & SEIR 2,3 &  & SI & \\
Init/Prob & Real & Binary & Real & Binary & Real & Binary & Real & Binary \\
\noalign{\smallskip}\hline
33/0.75 & 0.024 & 1 & 0.019 & 1 & 0.016 & 1 & 0.018 & 1 \\
33/0.50 & 0.040 & 1 & 0.033 & 1 & 0.03 & 1 & 0.02 & 1 \\
33/0.25 & 0.079 & 1 & 0.041 & 1 & 0.13 & 1 & 0.05 & 1 \\
\noalign{\smallskip}\hline
10/0.75 & 0.024 & 1 & 0.022 & 1 & 0.015 & 1 & 0.016 & 1 \\
10/0.50 & 0.054 & 1 & 0.053 & 1 & 0.043 & 1 & 0.027 & 1 \\
10/0.25 & 0.14 & 0.99 & 0.068 & 1 & 0.17 & 1 & 0.086 &  1 \\
\noalign{\smallskip}\hline
2/0.75 & 0.023 & 1 & 0.0181 & 1 & 0.016 & 1 & 0.015 & 1 \\
2/0.50 & 0.046 & 1 & 0.0355 & 1 & 0.0254  & 1 & 0.022 & 1 \\
2/0.25 & 0.073 & 1 & 0.14 & 0.99 & 0.0812 & 1 & 0.055 & 1 \\
\noalign{\smallskip}\hline
\end{tabular}

\end{table*}

The RMSE  and the AUC values are presented for the real-valued and binary infection tasks, respectively, for the two-observation case ($t_0 = i_0$, $t_1 = t_{max} = i_{max}$)
 in Table 2. Results can be seen for all four of the infection models and all possible initial infection states, and infection probability combinations. In the first column, the first number denotes the
percentage of initially infected nodes, the second one the maximum values of the edge infection probabilities.
In the SI model the second observation was selected so that the estimated averaged node infection value at the observation was close to the estimated averaged node infection value
at the last of observation of the SEIR model.
As we saw in the previous section, the binary tasks are much easier to compute than the real-valued scenarios. 
Based on the results in Table 2, the larger infections are easier to predict. The reason for this is that the infection
probability of a node is bounded: no matter how dense the infections are, the node infection probability cannot be greater than one. This means, that if the infections are dense --
either because there are many initially infected nodes, or the edge infection probabilities are large -- a sizable fraction of the nodes are going to have an infection value close to one.
This makes their distribution more homogeneous and this also makes the optimization task easier, since there are less distinct values to estimate.

The difference between the infection models is deceptive. Since we are running the same infection tasks on the same network with different infection models, if we increase the
infectious period we can expect greater prevalence of vertex infections, therefore we are seeing the same phenomenon as with the size of infections.
To confirm this we have shown the average infection probability of the nodes for the different tasks in Table 3, and indeed they correlate strongly with the error seen in Table 2.

\begin{table}[!t]
\caption{Average node infection value for the last observation for all infection models. As before, the numbers of the inverse infection tasks correspond to the scenarios described in this section. The first number denotes the percentage of initially infected nodes, the second one the maximum values of the edge infection probabilities.}
\centering
\begin{tabular}{lllll}
\hline\noalign{\smallskip}
Task & IC & SIR 2 & SEIR 2,3 & SI \\
Init/Prob &  &   &  &  \\
\noalign{\smallskip}\hline
33/0.75 & 0.78 & 0.87 & 0.91 & 0.9 \\
33/0.50 & 0.65 & 0.8 & 0.86 &  0.88  \\
33/0.25 & 0.46 & 0.65 & 0.72 & 0.72 \\
\noalign{\smallskip}\hline
10/0.75 & 0.69 & 0.83 & 0.88 & 0.85 \\
10/0.50 & 0.51 & 0.72 & 0.8 & 0.79 \\
10/0.25 & 0.22 & 0.46 & 0.6 & 0.56 \\
\noalign{\smallskip}\hline
2/0.75 & 0.68 & 0.82 & 0.88 & 0.85 \\
2/0.50 & 0.50 & 0.72 & 0.8 & 0.79 \\
2/0.25 & 0.20 & 0.45 & 0.6 & 0.56 \\
\noalign{\smallskip}\hline
\end{tabular}
\end{table}

\begin{figure*}[!t]
\label{fig:2}
\centering
\includegraphics[scale=0.8]{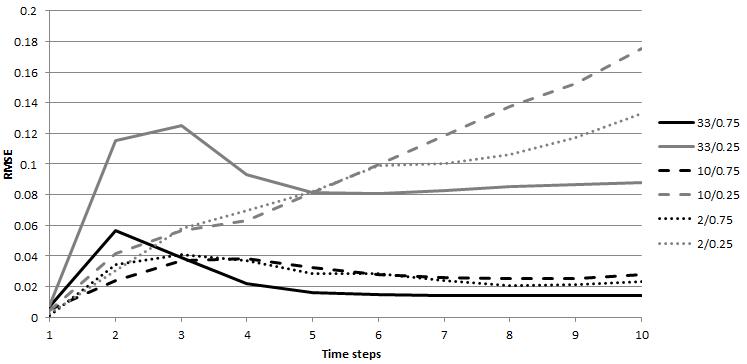}
\caption{RMSE at different time steps of the infection process for the fully observed scenario in the SIR model. The error is shown on the observations at $T : \{ 1 \dots 10\}$}
\end{figure*}

Results for the fully observed real-valued scenarios are shown in Figure 4.
The error is computed on the observations at each time step, $T : \{ 1 \dots 10\}$, for the SIR model. All infection scenarios were run for 10 iterations. The same performance behavior was observed for the other infection models, and they are, therefore, not presented separately. 
Figure 4 illustrates the significant difference in the behavior of the inverse infection tasks with sparse and dense infections. When the infections are large, in the first few observations the
error increases, then steadily decreases to its final value. With sparse infections the situation is more similar to what we observed in trees; the error continuously
increases until the end of the infection process. The tasks with dense initial
infections and sparse infection spreading behaves somewhat like a combination of the two.
The explanation behind this behavior is the following.
If the initially infected nodes are far apart from each other, the first few steps of the process behave very much like what we have seen on the trees of the previous section, and the reasoning remains the same. Furthermore, if the infection probabilities are small, the infections stay local, centered around the initially infected nodes. However, if the infections originating from multiple sources reach each other, the probability of being infected rises significantly due to multiple independent infection paths connecting each node. However, if the infections are dense enough, again a sizable fraction of the nodes will have an infection value close to one, making the optimization
task easier, and explaining the improvement in model performance seen on Figure 4 for more aggressive infection processes.

Finally, from the optimization perspective the task is moderately difficult for the FPSO algorithm. The performance varies depending on the task: the binary ones are easy, always taking less than two hundred iterations to conclude, while the real-valued scenarios with sparse infection are the most difficult often requiring over a thousand iterations. The accuracy of the predictions is acceptable; for the majority of the test runs the error stays below 5\%, but there are cases where the search is less successful, for example the
10/0.25 scenario for the IC model. Nonetheless we can conclude the the optimization method performs well in this application.

\subsection{Scalability}

We close the evaluation by investigating the scalability of the optimization method. Our interests here are the average number of iterations the method requires to produce
a solution, the total running time\footnote{A parallel version of the algorithm was implemented in C++, and the results were computed on a PC with an 4-core i7-4770 3.4 GHz processor.},
and the relationship between the size of the task and the accuracy of the estimation.


We examine four graphs in this section with $1000, \dots, 4000$ nodes and $1754, 3752, 5700, 7753$ edges and four infection scenarios from the previous section: 33\% or 10\% of the
nodes are initially infected with the infection probability between 0 and 0.5 for the real-valued case and uniformly 1 for the binary case. Reference edge infection values were drawn from
an uniform random distribution between 0 and 0.25, 0 and 0.75 respectively. Both binary and real-valued two-observation infection scenarios are investigated. We only show results for the
SIR $\tau_i = 2$ model, the performance of the other models was similar. Each of these tasks was computed ten times and
the results were averaged. The variance of the results was minimal. 

\begin{figure*}[!t]
\label{fig:3}
\centering
\includegraphics[scale=0.75]{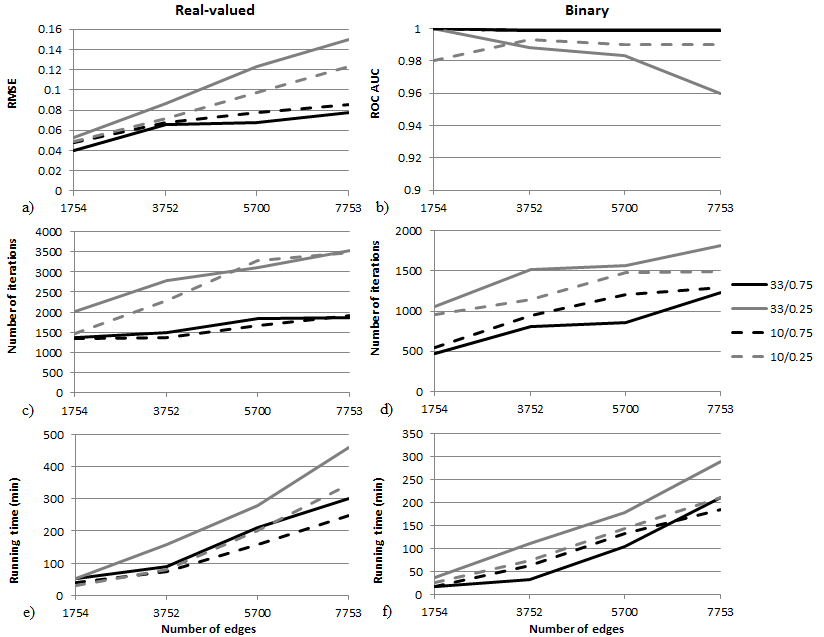}
\caption{Scalability results for the two-obervation infection scenarios on four graphs. Both binary and real-valued cases are shown. RMSE and ROC AUC values can be seen on a) and b),
average number of iterations on c) and d), and the averaged total running time in minutes on e) and f).}
\end{figure*}

Figure 5 shows the accuracy of the estimations, the averaged number of iterations and running time for the above infection scenarios. The binary tasks remain much 
easier to solve than the real-valued tasks, requiring fewer iterations and less computational time to find a good estimation. The difference between the infection scenarios is also similar to that observed in the smaller examples, with larger infections being easier to predict. Although, this depends on the ability of the graph to transfer infections (the reference edge values), as opposed to the fraction of 
initially infected vertices.

Accuracy scaling can be seen on Figures 5/a and 5/b. The error of the estimation increases with the number of estimated values (edges), although the increase is significantly lower if the
reference edge values are large. RMSE error values stay below $0.09$ for all graphs if edge infections are large, but they are greater than $0.1$ if infection values are small. Similar
behavior can be observed in the binary case, where the classification is perfect if the transfer probabilities are large, but stays below $1$ if the edge values are small.
The difference between scenarios with large and small edge infection probabilities can also be seen on Figures 5/c and 5/d, where the number of iterations required to find a good
solution is much smaller if edge infection values are large. Finally, the running time of the tasks in minutes can be seen on Figures 5/e and 5/f.
Depending on the size of the graph and the edge infection values the tasks took one to eight hours in average to compute on the computer listed above.

Directly estimating the edge infection values in GIIM is a difficult task, since the number of values to estimate \textit{i.e}. the number of edges, can be large even for medium sized graphs, posing
 a challenge for the optimization algorithm. In this section we considered graphs up to $4000$ nodes and $7753$ edges, and seen that the estimation is possible on an average PC with acceptable accuracy and running time. 

There are multiple ways to speed up the method so that it can handle large networks. 
The time complexity of the optimization depends heavily on the
time complexity of the infection model. A modified version of the CompleteSim infection heuristic \cite{acta} was used in this paper, but other potentially faster heuristics are available
in the literature \cite{acta,kinai2}. Another way to considerably speed up estimations is to assume that the edge infection values can be computed as functions of known vertex or
edge attributes. In this case only the coefficients of these functions have to be estimated, which are far less numerous than the number of edges in the graph. 

\section{Future works}

The scope of this paper allows us to test only part of the functionality provided by the Generalized Inverse Infection Model. In the previous sections we presented the behavior of GIIM with respect to the number of type of available observations, infection model and the specific infection scenario. However, there are many additional applications and extensions of the proposed model. Certain extensions are discussed below, and will be explored in future works.

\subsection{Function estimation}

The task of GIIM is to find an edge weight assignment that minimizes the difference between the reference and computed infections. This requires finding the edge infection
probability for all edges of the graph. Optimizing for this many values, while possible as we have seen in Section 4, is difficult even for medium sized graphs.

Another approach to direct edge weight estimation is to compute the edge weights as a function of existing attributes as seen in \cite{iip,cejor,lg2}. 
These works assume that additional information can be assigned to the edges or vertices of the graph. Examples include the number of passengers or travel distance in an air traffic network \cite{lg2}, and the amount or frequency of transactions in \cite{cejor}. The edge infection probabilities themselves can be defined as functions of these attributes, in which case the optimization task changes to finding the proper function formulation explaining the reference infection process. This approach is often an easier estimation task compared with finding all of the edge weights. 

It is possible to extend the notion of formulating edge infection values as functions of edge attributes to vertices. If observations are available on nodes that relate to the infection
process it is possible to define a function that computes a single infection value for the node. Then these values can be used as observations. 
The values might be normalized between zero and one, although it is important to note that these are generally not probability values.


\subsection{Time scale of the infection process}

So far we have assumed that the sample times of the reference and computed observations are the same: $T = T'$ in $O = Inf(G, W,  \mathcal{I}, T)$, $O' = Inf(G, W',  \mathcal{I}, T')$.
This is usually not realistic, since many times we cannot freely pick the exact moment the observations are made or the amount of observations. Real-life sample times are continuous, for example, dates or clock time. Therefore, matching them to the sample times of the simulated observation process may be non-trivial especially if they are noisy as well.
Additionally, infections in the real world might not behave the way we expect them to. They can be influenced by factors unknown to the observer, suddenly become more virulent or stop for no
apparent reason: the time scale of the infection process might be unknown. The relationship between the timescale of the observed and computed processes might not be linear or not even
monotonic. Therefore, matching the observations on an artificially simulated infection process to real-life observations might be a difficult task.


\subsection{Identifying and calibrating the infection model}

Another point of interest is the epidemic model. In some real-life applications it is fairly possible that 1, the infection model is unknown 2, the parameters of the infection model are unknown.
In the first case it might be worthwhile to start with a more general model and move to more specific ones. The second case is simpler, instead of looking for $W'$ in
$O' = Inf(G, W',  \mathcal{I}(\theta'), T')$ we are looking for $\theta'$ or possibly both, where $\theta'$ denotes the parameters of the infection function. If the edge weights are unknown,
this might be a difficult task, since we are optimizing for both $W'$ and $\theta'$.

\subsection{Planned application}

The next phase of this research will be the application of the GIIM to a global epidemic. Observing the spread of infectious diseases is not an easy task. However, there are
examples of well-documented outbreaks. Specifically, the estimated number of infections over time in either a given city, region and country are increasingly available for large scale outbreaks. For international outbreaks, where the transfer method between the countries can be attributed to a specific medium: air or maritime transportation, the GIIM can be applied to estimate the risk of disease spread between regions, via specific travel routes. This type of analysis can aid in both prediction, prevention and disease control, and will be explored in the context of the 2015 Zika pandemic in future work.

\section{Conclusions}

In this paper we proposed the Generalized Inverse Infection Model with the intention of providing a general framework to describe inverse infection tasks. Given an unweighted graph
and a set of observations on an infection process GIIM is able to estimate the edge infection probabilities explaining the input observations. 
Contrary to existing models our model is able to handle an arbitrary number of these; it can be applied to fully, minimally or potentially partially observed infection processes.
The nature of the observations is not fixed; in the scope of this paper we experimented with binary and real node infection values but it is possible to use other values tied to
the spread of the infection process within the framework. The underlying infection model may also be customized: we worked with the SI, IC, SIR and SEIR models but it is also
possible to consider other network based models. GIIM formulates the inverse infection task as a general optimization problem and a Particle Swarm Optimization method was suggested as
the solution method. Finally, it is possible to further extend the functionality introduced in this paper, like estimating functions on the edges instead of 
probability values or estimating the parameters of the infection model. Additional suggestions can be found in Section 5.

We evaluated the performance of GIIM on a high variety of artificial infection scenarios. 
Our goal was to investigate the relationship between the parameters of the infection scenario and
the difficulty of the edge estimation. We have shown on trees that even though the problem is underdetermined it is possible to find reasonably accurate explanations of the infection
process. On more general complex networks we examined the behavior of our method on 108 different inverse infection tasks. The tasks varied according to the epidemic model, the
density of initial infections, the reference edge infection probabilities and the type and number of observations. Our findings indicate, that there is close relationship between the expected
size of the infection and difficulty of the task, with larger infections being easier to predict. There is complex interplay between density of initial infections, the likelihood of the infection spread
and the difficulty of the estimation at different time steps. We can also conclude that even though (or perhaps because of) binary tasks are more underdetermined, they are much more
easy to predict. Finally we have experimented with larger networks and a few specific infection scenarios to investigate the scalability of the problem. We can say, that the optimization
method performs well for almost all of the situations described in this paper.

The framework of GIIM  allows the estimation of edge infection probabilities in scenarios different from the ones we have seen in the evaluations section.
We provided a couple of suggestions to further improve the functionality described in this paper as well as a planned application. 

\bigskip

\noindent{\bf Acknowledgements}

We thank the National Health and Medical Research Council (NHMRC) for funding, project grant (No. APP1082524). The contents of the published material are solely the responsibility of the Administering Institution, a Participating Institution or individual authors and do not reflect the views of the NHMRC.



\end{document}